\documentclass[fleqn,10pt,twocolumn]{SICE_FES26}

\title{A Wearable Sensor Array on Conductive Textiles\\
for Body Surface Temperature Mapping}

\author{Amiri Onuki${}^{1}$ and Akihito Noda${}^{1}$}

\affils{${}^{1}$Kochi University of Technology, Kochi, Japan\\
(E-mail: 305036n@gs.kochi-tech.ac.jp, noda.akihito@kochi-tech.ac.jp)\\
}

\abstract{%
Continuous monitoring of body-surface temperature distributions can provide useful information for wearable health and thermal-condition monitoring. However, practical daily-use wearable systems for distributed temperature sensing remain difficult to implement because multiple sensor nodes require power delivery and data collection across a garment. Conventional wiring increases system complexity, reduces placement flexibility, and can compromise the wearability of garment-integrated sensor arrays. To address this issue, we present a conductive-textile-based wearable temperature sensing system in which the textile itself serves as a shared bus for both DC power delivery and data communication. Each sensor node is implemented as a compact, battery-free circular board with a diameter of 21 mm and a weight of 0.9 g, integrating a commercial temperature sensor and custom circuitry for UART/DC superposition.
The shared textile bus enables flexible node placement without individual wiring while supplying power to each node.
Address-based half-duplex UART with master polling enables collision-free multi-node communication. To demonstrate the system-level capability, we implemented an 18-node sensor array on a garment worn by a mannequin and developed a real-time temperature mapping viewer. In a thermal scanning demonstration using warm air from a heat gun with a set temperature of approximately 50°C, the 18-node array captured spatiotemporal temperature variations over the garment surface at a refresh rate of 99.2 Hz. These results show that the proposed conductive-textile bus can support real-time, high-refresh-rate distributed temperature mapping, providing a practical basis for garment-integrated body-surface monitoring and future multi-sensor wearable systems.
}


\usepackage{makecell}
\usepackage{url}

\def\finaldoi{}
\def\noticedoi{\ifx\finaldoi\empty
  The final published version is available via IEEE Xplore.%
  \else
  The final published version is available at
  \texttt{https://doi.org/\finaldoi}.%
  \fi}
\newcommand{\copyrightnotice}{%
\textcopyright~2026 The Society of Instrument and Control Engineers (SICE).
This is the authors' accepted manuscript of a paper published in
\textit{Proceedings of the SICE Festival 2026 with Annual Conference},
September 14--17, 2026, Yokohama, Japan. \noticedoi{}
Personal use of this material is permitted.
Permission from SICE must be obtained for all other uses, including
reprinting/republishing, sale, use for advertising or promotional purposes,
and reuse of any copyrighted component of this work in other works.}

\makeatletter
\def\ps@titlepagestyle{%
  \def\@oddhead{}\def\@evenhead{}%
  \def\@oddfoot{\parbox[t]{\textwidth}{\vspace*{-6.5mm}%
    \scriptsize\raggedright\copyrightnotice}}%
  \let\@evenfoot\@oddfoot}
\def\maketitle{\par
  \begingroup
  \twocolumn[\@maketitle]
  \thispagestyle{titlepagestyle}%
  \endgroup
  \setcounter{footnote}{0}\let\maketitle\relax\let\@maketitle\relax
  \gdef\@thanks{}\gdef\@author{}\gdef\@title{}\gdef\@abstract{}
  \gdef\@keywords{}\gdef\@affils{}\let\thanks\relax
}
\makeatother

\begin{document}

\maketitle

\section{Introduction}

Wearable measurement technologies enable continuous monitoring of physiological signals in daily life \cite{lifelog}. 
In addition to single-point measurements, spatially distributed sensing can provide information about local variations over the body surface. 
Body-surface temperature is affected by physiological and environmental factors, such as peripheral blood flow, thermo regulation, inflammation, ambient temperature, airflow, and clothing conditions \cite{temperature}. 
Mapping its spatial distribution can therefore help capture localized thermal changes that are difficult to observe from a single measurement point.

Existing approaches for body-surface temperature monitoring include infrared thermography, wrist-worn wearable devices, and skin-attachable sensors. 
Infrared thermography can visualize wide-area temperature distributions without contact \cite{temperature,ThermalCamera}, but it requires an external camera and is not suited to continuous daily monitoring. 
Wrist-worn wearable devices provide a practical solution for long-term point measurements with a low user burden \cite{Smartwatch}, whereas skin-attachable sensors can measure body-surface temperature more directly and can be extended to distributed temperature mapping \cite{skin,ex}. 
However, wide-area temperature mapping over the body requires multiple distributed sensing points, which introduces additional challenges in power delivery, data collection, and wearability.

Skin-attachable sensors can measure body-surface temperature directly by maintaining close contact with the skin. 
One representative approach uses battery-free wireless sensor nodes and near-field communication (NFC) for full-body pressure and temperature mapping \cite{ex}. 
This approach is well suited to stationary monitoring, where passive sensor nodes can be read through an external primary antenna embedded in the surrounding environment, such as under a hospital bed. 
However, such an environment-assisted architecture is not fully self-contained on the wearer. 
The reported wireless power transfer of 4--12~W and the scanning time of approximately 3~s for 65 sensors also indicate a different design tradeoff from high-refresh-rate, low-power temperature mapping on clothing.

Clothing-integrated sensing provides a practical form factor for covering large areas of the body while maintaining wearability during daily activities. 
However, distributed sensor arrays on clothing require power delivery and data collection across multiple locations on a deformable fabric. 
Individual wiring increases system complexity, restricts sensor placement, and can reduce connection reliability under garment deformation and repeated body movements \cite{wire}. 
Battery-powered wireless sensor nodes reduce wiring but increase node size, weight, power-management complexity, and maintenance burden \cite{WIFI}. 
These issues motivate a shared infrastructure for supplying power and collecting data from multiple sensor nodes while preserving garment flexibility and wearability.

Table~\ref{tab:comparison} compares representative approaches for distributed wearable sensing. 
\begin{table*}[t]
\centering
\caption{Representative implementation strategies for wearable sensor arrays.}
\label{tab:comparison}
\small
\renewcommand{\arraystretch}{1.2}
\begin{tabular}{lccccc}
\hline
\makecell{\textbf{Implementation}\\\textbf{type}} &
\makecell{\textbf{Battery-free}\\\textbf{node}} &
\makecell{\textbf{Fully}\\\textbf{wearable}} &
\makecell{\textbf{Placement}\\\textbf{freedom}} &
\makecell{\textbf{Wiring}\\\textbf{robustness}} &
\makecell{\textbf{Typical}\\\textbf{use}} \\
\hline
\makecell[l]{Individually wired sensors \cite{wire}} &
$\bigcirc$ & $\triangle$ & $\triangle$ & $\times$ &
Lab / validation \\

\makecell[l]{Wi-Fi/Bluetooth sensor nodes \cite{WIFI}} &
$\times$ & $\bigcirc$ & $\bigcirc$ & $\bigcirc$ &
Daily monitoring \\

\makecell[l]{Battery-free NFC/RFID sensors \cite{ex}} &
$\bigcirc$ & $\times$ & $\bigcirc$ & $\bigcirc$ &
Bed / proximity \\

\makecell[l]{Conductive-textile-connected sensors (this work)} &
$\bigcirc$ & $\bigcirc$ & $\bigcirc$ & $\triangle$--$\bigcirc$ &
Garment sensing \\
\hline
\end{tabular}

\vspace{1mm}
\footnotesize{
$\bigcirc$: favorable, 
$\triangle$: partially limited, 
$\times$: unfavorable.
}
\end{table*}
Individually wired sensors provide straightforward power delivery and data collection, but wiring complexity and connection points increase with the number of sensors. 
Battery-powered wireless nodes reduce wiring but require a battery, wireless circuitry, and power management at each node. 
Passive NFC/RFID-based sensors eliminate on-node batteries but generally rely on nearby external reader antennas, making the sensing area dependent on the reader placement. 
These limitations motivate a garment-integrated shared infrastructure that can supply power and collect data from multiple distributed sensor nodes.

To address this need, this study uses conductive textiles as a shared bus for distributed temperature sensing on clothing. 
Digital communication signals are superimposed on the DC power line, allowing each node to receive power and exchange data through the same textile medium. 
This architecture eliminates individual wiring, on-board batteries, and wireless antennas at each sensor node while preserving flexible node placement on the garment.

The contributions of this paper are threefold. 
First, we present a conductive-textile-based shared-bus architecture for battery-free, distributed temperature sensing on clothing. 
Second, we implement compact sensor nodes with a diameter of 21 mm and a weight of 0.9 g, integrating a commercial temperature sensor and custom circuitry for UART/DC superposition. 
Third, we demonstrate an 18-node garment-integrated prototype with a real-time temperature mapping viewer, achieving a refresh rate of 99.2 Hz. 
These results indicate the feasibility of real-time, high-refresh-rate temperature mapping on clothing and provide a basis for future high-resolution body-surface monitoring using garment-integrated sensor networks.

\section{System architecture and principles}

\begin{figure}[t]
    \centering
    \includegraphics[angle=90, width=1.0\linewidth]{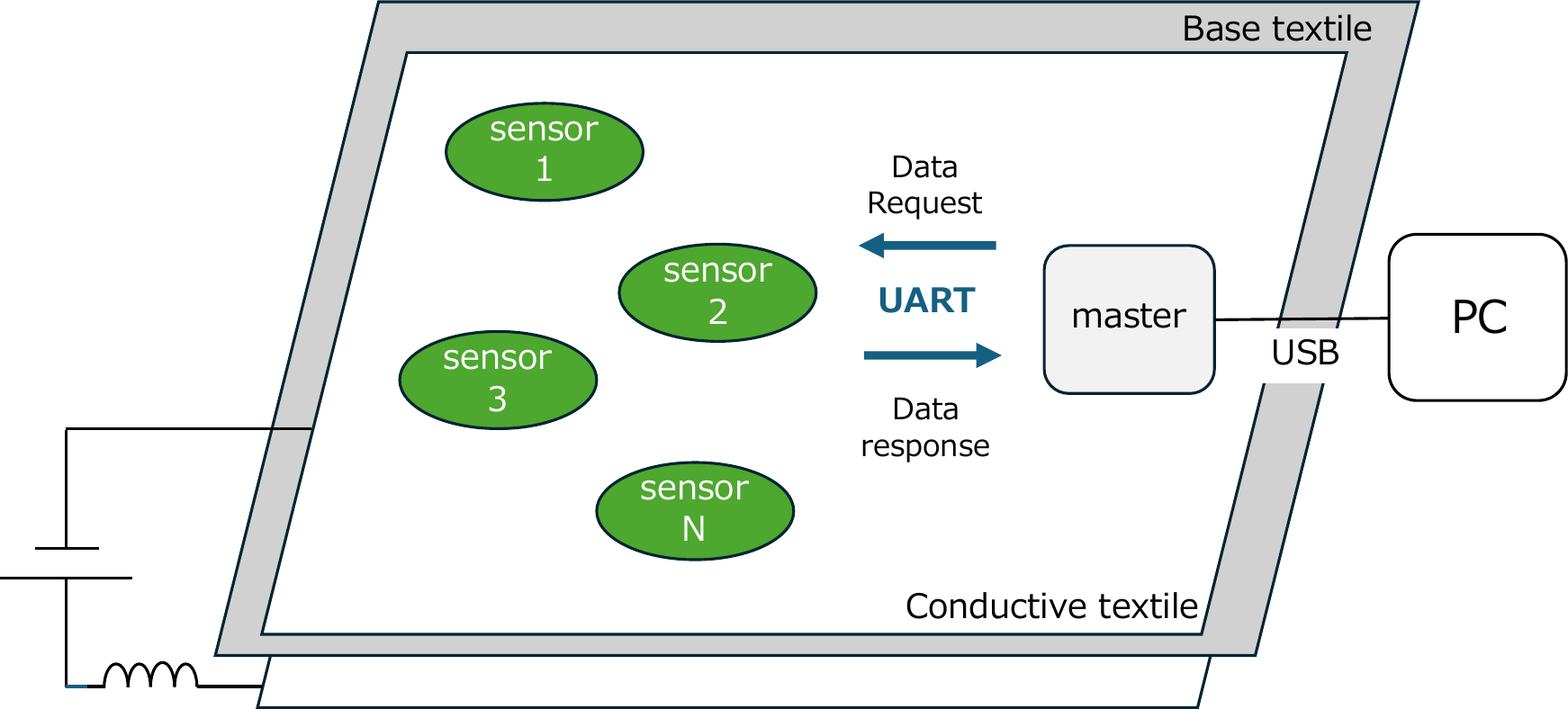}
    \caption{Overall system configuration. The master node controls network communication, aggregating data from all sensor nodes for display on a PC.}
    \label{fig:system}
\end{figure}

The overall configuration of the proposed system is shown in Fig.~\ref{fig:system}.
The textile transmission medium consists of two conductive textile layers separated by an insulating fabric layer and functions as a shared transmission line for the entire system.
Each sensor receives power through this conductive textile and transmits the acquired temperature data to the master node.
The master node coordinates the communication of the entire system, aggregates multi point data, and visualized it as a real-time heatmap on a PC.

To simultaneously achieve DC power supply and bidirectional data communication over a single transmission line, the system employs the following communication method proposed in our previous and ongoing work.

The proposed system utilizes two-dimensional communication on conductive textiles \cite{ConductiveTextile}.
The transmission medium features a three layered structure consisting of two conductive textile layered (VCC and GND layers) with an insulating fabric layer sandwiched between them.
Devices are fixed to arbitrary positions on the textile and establish electrical connections simultaneously via pin badge shaped connectors.
Through this architecture, the entire conductive textile functions as a common power and signal transmission line, allowing all sensors on the fabric to be connected electrically in parallel.
As a result, this architecture eliminates the need for individual wiring and separate batteries for each device.

To supply DC power and perform data communication simultaneously over a single transmission line, this system adopts a superimposed transmission method.
When a transmitter outputs a digital UART signal, it is superimposed as spike-like voltage variations onto the DC power supply layer (VCC) at the rising and falling edges of the signal.
On the receiving end, these analog variations on the conductive textile are evaluated against predefined voltage thresholds.
This process effectively demodulates the signals back into the original digital waveforms recognizable by the microcontroller unit (MCU).

While I$^2$C communication is employed in existing conductive textile networks, 
it faces several challenges \cite{I2C}.
These include difficulties in multi-drop connections for identical devices, 
increased circuit complexity due to the requirement for two different carrier frequencies, and significant constraints on communication speed.
To address these issues, the proposed system adopts UART using baseband transmission, which does not require a carrier wave \cite{U4We}.
This approach simplifies the physical layer by eliminating the need for complex modulation and demodulation hardware.
Furthermore, to achieve bidirectional communication over a signal line, the system implements half-duplex communication.
The transmission and reception drivers are dynamically switched via a control signal from the MCU.
This configuration ensures robust data exchange while minimizing the physical footprint of the sensor nodes on the textile.

\section{Integrated Temperature Sensor Node}
The successful implementation of the proposed conductive textile network requires sensor nodes capable of simultaneous power harvesting and bidirectional data transmission in a highly compact form factor. 
To meet these specific hardware requirements, we developed a completely integrated, custom sensor node. 
Its circuitry is primarily divided into two functional blocks:
the temperature measurement and control unit, and the communication and signal processing unit.

\subsection{Temperature Measurement and Control Unit}

It is equipped with a digital temperature sensor (TMP117, Texas Instruments) and a microcontroller (MCU: ATmega328P, Microchip Technology).
The TMP117 provides a resolution of 0.0078~$^\circ$C,
which is sufficient for capturing subtle physiological change on body surface.
The MCU retrieves the temperature data from the sensor via an I$^2$C interface and prepares it for transmission.

\subsection{Communication and Signal Processing Unit}
This unit physically realizes the transmission principles described in Section II.
It primarily consists of the following three hardware elements:

\begin{itemize}

\item \textbf{Power and Signal Separation:}
To separate the superimposed DC power and high-frequency UART signals, an LC filter is employed. 
The inductor isolates stable DC power for the MCU, while coupling capacitors extract the AC communication signals.

\item\textbf{Half-Duplex Switching:}
Unlike previously reported receive-only modules, this sensor node requires bidirectional communication to realize the polling-response architecture.
To implement half-duplex communication over a single shared bus, an analog switch IC is employed.
This switch is dynamically controlled by the Tx\_EN signal from the MCU.
From a hardware perspective, strict state control is essential to prevent signal collisions.
The MCU asserts Tx\_EN to High, waits for a hardware settling margin to stabilize the circuit, transmits the UART frame, and immediately deasserts Tx\_EN to Low to return to the receiving state.
The detailed time margins and the corresponding communication sequence are discussed in Section 4.

\item\textbf{Signal Demodulation:}
To demodulate the spike waveforms into digital signals, an op-amp-based hysteresis comparator is employed. 
With thresholds defined by two resistors, the comparator shapes minute analog variations on the conductive textile into clear rectangular pulses for the MCU.

\end{itemize}

\begin{figure}[t]
    \centering
    \includegraphics[angle=90, width=1.0\linewidth]{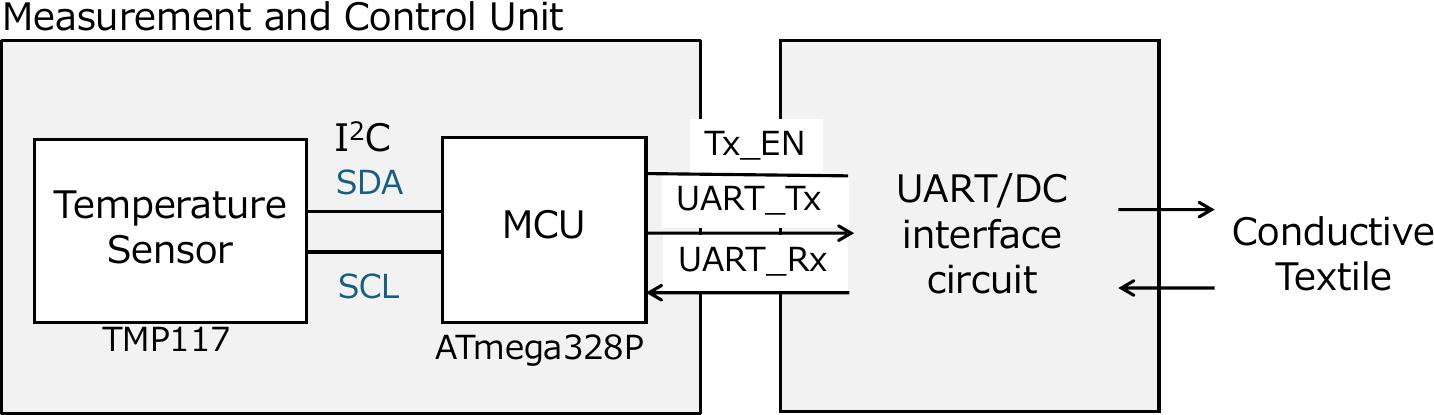}
    \caption{Block diagram of the sensor node. The MCU acquires temperature data via I$^2$C and transmits to the textile bus through the UART/DC interface circuit.}
    \label{fig:kairozu1}
\end{figure}

\subsection{PCB Implementation}

\begin{figure}
    \centering
    \includegraphics[width=1.0\linewidth]{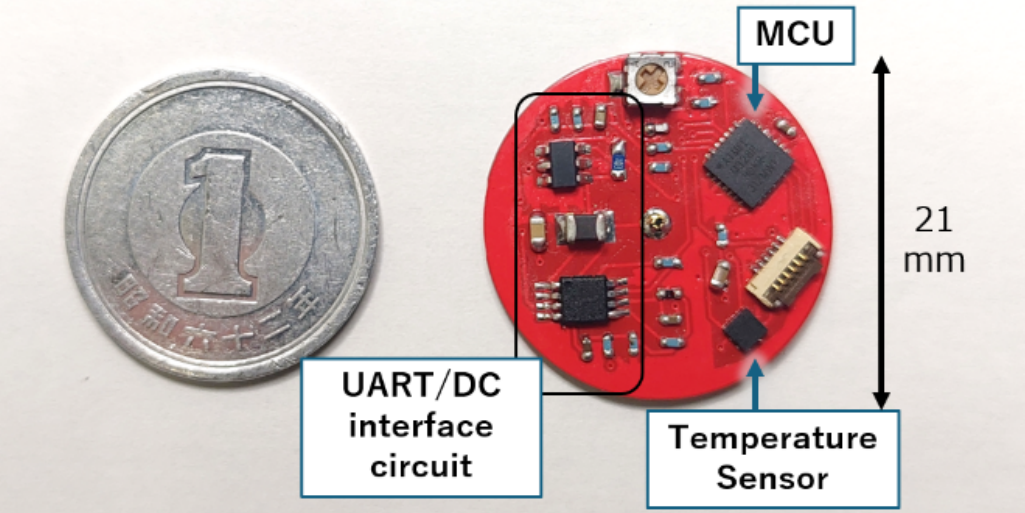}
    \caption{Implemented compact circular PCB.}
    \label{fig:kiban}
\end{figure}

Fig.~\ref{fig:kiban} shows the compact, circular PCB that integrates these two circuits.
It has a diameter of 21~mm and a weight of 0.9~g.

To prevent the board's internal heat generation from affecting the measurements, the temperature sensor is placed at the extreme edge of the PCB.
Furthermore, managing the power consumption of the compact sensor board is a critical practical consideration. 
The integrated node consumes approximately 110~mW during active communication. 
Given the small 21~mm footprint, continuous active operation leads to non-negligible self-heating of the MCU and circuits, which can conductively affect the TMP117 sensor readings. 
Therefore, mitigating this thermal influence through communication control is essential.

\section{Communication Protocol}
\subsection{Addressing and Collision Avoidance}
Standard UART is designed for point-to-point connections. 
To enable multi-node communication on a shared bus without signal collisions, we assigned a unique address ID to each sensor and implemented a master-polling architecture. 
The master node retains exclusive authority to initiate communication, sequentially polling each node to complete one full measurement cycle.
Meanwhile, sensor nodes remain passive, transmitting data only upon receiving a request matching their specific ID. This architecture effectively eliminates data collisions on the shared conductive textile.

\subsection{Communication Flow}
To realize the half-duplex master-polling architecture, the transaction between the master and sensor nodes follows a strictly defined sequential flow controlled by the Tx\_EN signal:

\begin{enumerate}
\item \textbf{Request Transmission:}
The master node switches its Tx\_EN signal to High, transmits a request frame containing the target address, and immediately returns Tx\_EN to Low to enter the receiving state.

\item \textbf{Address Verification:} 
All sensor nodes constantly monitor the bus. 
Upon receiving the request, each node verifies the target ID. If the ID does not match its own, the node ignores the frame and maintains its receiving state.

\item \textbf{Data Response:} 
Only the sensor with the matching ID acquires the temperature data, switches its Tx\_EN to High to transmit the response frame, and immediately returns Tx\_EN to Low after the transmission is completed.

\item \textbf{Data Collection:} 
The master receives the response, transfers the data to the PC, and proceeds to poll the next sensor ID in the sequence.
\end{enumerate}

\subsection{Data Format}

\begin{figure}[t]
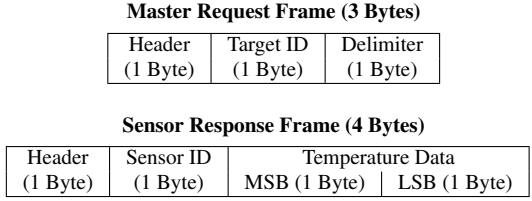

    \centering
    {\small
    \textbf{Master Request Frame (3 Bytes)}\\[1mm]
    \begin{tabular}{|c|c|c|}
        \hline
        Header & Target ID & Delimiter \\
        (1 Byte) & (1 Byte) & (1 Byte) \\
        \hline
    \end{tabular}
    
    \vspace{4mm}
    
    \textbf{Sensor Response Frame (4 Bytes)}\\[1mm]
    \begin{tabular}{|c|c|c|c|}
        \hline
        Header & Sensor ID & \multicolumn{2}{c|}{Temperature Data} \\
        (1 Byte) & (1 Byte) & MSB (1 Byte) & LSB (1 Byte) \\
        \hline
    \end{tabular}
    }
    \vspace{2mm}
    \caption{Data frame structures for the polling-response communication.}
    \label{fig:data_format}
\end{figure}

To minimize transmission overhead, data is exchanged in a fixed-length binary format rather than ASCII strings. 
The structures of the master's request frame (3 bytes) and the sensor's response frame (4 bytes) are illustrated in Fig.~\ref{fig:data_format}. 
This optimization reduces the payload size per transaction to only 7 bytes in total, which is essential for maximizing the overall sampling frequency and reducing the bus occupancy of the sensor network.

\subsection{Timing Analysis and Bus Occupancy}
\begin{figure}[t]
    \centering
    \includegraphics[angle=90, width=1.0\linewidth]{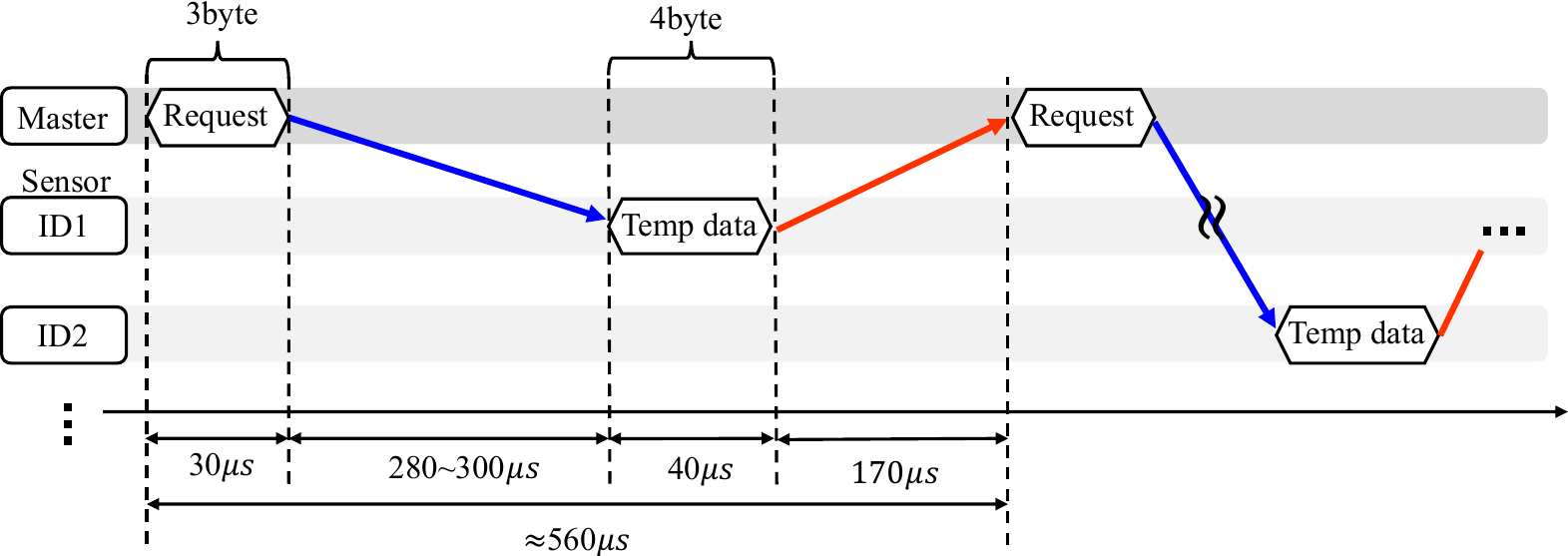}
    \caption{Timing sequence diagram of the polling-response communication.}
    \label{fig:polling}
\end{figure}

Operating at a UART baud rate of 1 Mbps, the theoretical communication time for a single polling cycle (7 bytes total) is approximately $70~\mu$s, as illustrated in the timing sequence in Fig.~\ref{fig:polling}.
In the current implementation, the master polls each node at an interval of approximately $560~\mu$s, resulting in a total array refresh rate of $99.2$~Hz (about 10 ms per cycle for 18 nodes).
However, for practical physiological applications, body surface temperature is a typical slow-varying signal that does not fluctuate dynamically within milliseconds~\cite{BodySurface}. 
Therefore, a measurement frequency of 1 Hz is generally more than sufficient to capture physiological thermal responses.

Critically, the actual data transmission ($70~\mu$s) occupies $12.5$\% of each $560~\mu$s polling interval. 
This low bus occupancy demonstrates two key advantages. 
First, the system has significant scalability to integrate additional high-speed sensors without reaching the physical limits of the communication line. 
Second, it provides a strategic margin for thermal mitigation.
For physiological applications where a 1 Hz refresh rate is sufficient, the system can place the sensor nodes in a low-power sleep state for the remaining $990$~ms of the 1-second interval. 
This drastically reduces the active duty cycle, suppressing the self-heating of the MCU and ensuring highly accurate physiological measurements.

\section{System Evaluation}
To validate the fundamental characteristics of the proposed sensor network before applying it to physiological measurements, we evaluated its scalability and thermal stability.

\subsection{Polling Rate and Scalability}
The experimental results demonstrated a strictly linear relationship, expressed as $560~\mu$s $\times N$. 
For instance, scanning half of the array ($N=9$) required approximately 5~ms, and the full array ($N=18$) required exactly 10~ms. 
This confirms that the system scales efficiently.

\subsection{Self-Heating Evaluation and Limitations}

The self-heating effect of the sensor nodes was evaluated in a $24~^{\circ}$C environment.
Starting from thermal equilibrium at $24~^{\circ}$C, the temperature reading reached $27~^{\circ}$C within 30~s after power-on, $30~^{\circ}$C after 2~min, and nearly saturated at $32~^{\circ}$C after 10~min.
This temperature rise was caused by the 110~mW power consumption of each sensor node.
In this demonstration, the sensor transmitted data at 99.2~samples/s, and the MCU remained continuously active to demonstrate high-rate sensing, resulting in the highest-power operating condition.
In practical use, this temperature rise can be reduced by lowering the sampling rate according to the required update frequency and putting the MCU into sleep mode between measurements.
The self-heating can be further reduced by replacing the MCU and op-amp with lower-power models.

\section{Demonstration of Temperature Mapping}

\begin{figure*}[t]
    \centering
    \includegraphics[angle=90, width=0.95\textwidth]{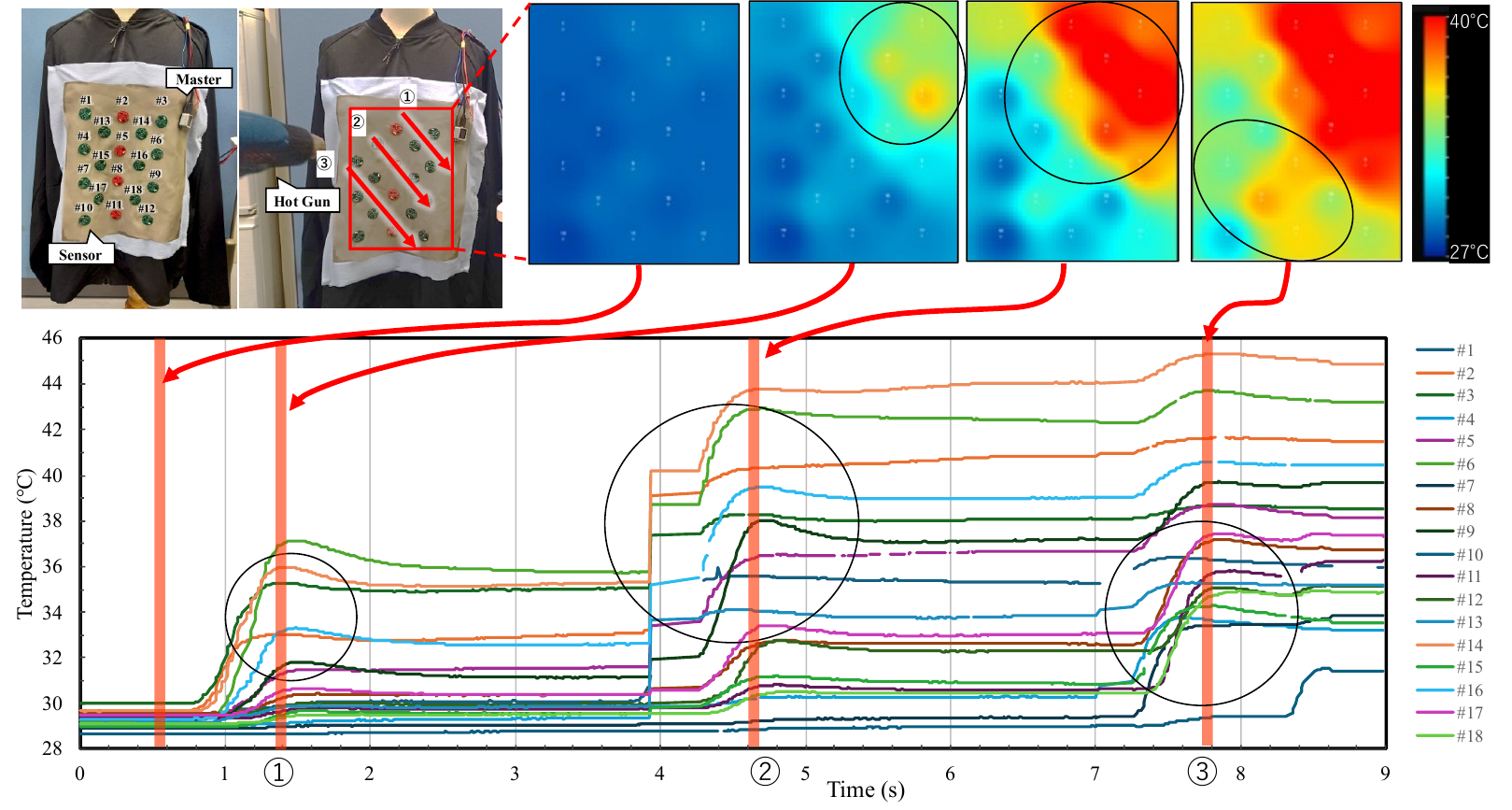}
    \caption{Experimental setup and real-time heatmap visualization of dynamic temperature mapping. 
    An 18-node sensor array was attached to a garment, and warm air from a heat gun was swept over the array along the indicated trajectory. 
    The heatmaps and temperature traces show the resulting spatiotemporal temperature variations over the garment surface.
    All data are raw sensor readings without self-heating compensation.}
    \label{fig:heat_tracking}
\end{figure*}

\subsection{Real-time visualization of temperature distribution}
To evaluate the effectiveness of the proposed system, 
we developed an application that generates real-time heatmaps using data acquired from multiple sensors distributed on the conductive textile.

To visualize the spatial temperature distribution, the system performs smooth interpolation between the discrete data points retrieved from the sensors.
Specifically, the estimated temperature $T(P)$ at an arbitrary point $P(x, y)$ within the display area is calculated as a weighted average of the temperatures $T_i$ measured by each sensor $i$, formulated as follows:

\begin{equation}
T(P) = \frac{\sum_{i=1}^{N} w_i T_i}{\sum_{i=1}^{N} w_i}
\end{equation}

where $N$ is the total number of sensors, and $w_i$ represents the weight coefficient based on the distance $d_i$ between point $P$ and the position of sensor $S_i(x_i, y_i)$.
In this system, the weight function is defined as:

\begin{equation}
w_i = \frac{1}{\left(\frac{d_i}{D}\right)^\alpha + \epsilon}
\end{equation}

where $D$ is a scaling constant that adjusts the spread of the temperature distribution, $\alpha$ is an exponent determining the strength of distance-based attenuation, and $\epsilon$ is a small constant to prevent division by zero when the distance is extremely small. 
Through this weighting scheme, the temperature of a specific sensor is strongly reflected in its immediate vicinity, while the regions between sensors represent a smoothly interpolated temperature distribution from multiple sensor values.

A setup with 18 sensor nodes arranged in a staggered pattern was attached to the surface of a garment.
Fig.~\ref{fig:heat_tracking} demonstrates the dynamic tracking capability of the system. Warm air from a heat gun was applied and swept across the sensor array along a specific trajectory.
The origin of the time axis corresponds to approximately 3~min after the system was powered on. The sensor nodes had almost reached thermal equilibrium, and the temperature readings remained nearly constant during the first 1~s in Fig.~\ref{fig:heat_tracking}.
Due to self-heating, the initial baseline ($t = 0$~s) was approximately 29--30~$^\circ$C under an ambient temperature of 24~$^\circ$C; i.e., a thermal offset of 5--6~$^\circ$C was included.
This experiment intended to demonstrate the temporal resolution and quick response of the 18-node sensor array implemented with proposed textile-based sernsor networking technique. 
Applying about 50~$^\circ$C warm air did not intend to evaluate the precision/accuracy required for biological applications. 
This confirms that the system's high refresh rate provides sufficient responsiveness to track dynamic and continuous thermal variations on the body surface.

\subsection{Limitations of the Current Prototype and Future Improvements}
Although the proposed system successfully demonstrated multi-node temperature monitoring, challenges remain for practical wearable deployment. 
First, scaling from the tested 18 nodes to hundreds of nodes requires further communication protocol optimization to manage data traffic without sacrificing refresh rates. 
Additionally, replacing the wired USB with a wireless master node (e.g., Wi-Fi/Bluetooth) is necessary for standalone operation. 
Second, robust operation in daily life demands improved wearability and noise tolerance. 
Since dynamic noise from body movements, textile deformation, and physical contact can destabilize communication waveforms, optimizing the physical characteristics of the conductive textiles and refining the mechanical connections are crucial future tasks.

Beyond network scalability and physical robustness, the measurement accuracy of the current sensor nodes themselves requires further improvement.
The sensor node presented in this work is an initial prototype, intentionally designed with a priority on ease of assembly and debugging rather than miniaturization and thermal optimization. 
Consequently, the current PCB has a relatively large thermal mass, which limits its transient thermal response. 
Furthermore, the MCU used in this prototype consumes relatively high power, and its self-heating degrades the temperature measurement accuracy at the sensing point.

It is crucial to note that these thermal limitations are not fundamental drawbacks of the proposed sensing concept, but rather artifacts of this specific implementation. 
The primary contribution of this work is demonstrating the feasibility of high-rate, distributed temperature sensing over conductive textiles.
Achieving a refresh rate exceeding 99~Hz with 18 nodes proves that the proposed communication architecture provides highly capable power and data transmission. 
The current thermal and accuracy issues can be definitively resolved through straightforward hardware optimizations in future iterations without losing this high-speed performance.

Specifically, replacing the TMP117 sensor with its wafer-level chip-scale package (WLCSP) variant would drastically reduce the thermal mass from 5.1~mJ/$^\circ$C to 0.8~mJ/$^\circ$C.
Similarly, adopting a modern, low-power MCU (e.g., STM32C011DxY) would significantly reduce self-heating, while its ultra-compact $1.7 \times 1.42$~mm footprint would further minimize the overall thermal capacity. 
Redesigning the PCB with a thinner substrate, minimized surface area, and optimized thermal isolation traces between the MCU and the sensor will effectively block internal heat conduction. 
In conclusion, the current results establish a strong proof of concept rather than an upper performance limit, and future versions will achieve substantially higher sensing accuracy and wearability while retaining the core advantages of the proposed communication scheme.

\section{Conclusion}
This study presented a conductive-textile-based wearable sensor network for body-surface temperature mapping. 
The textile serves as a shared bus for both DC power and UART communication, eliminating individual wiring and on-board batteries.
An 18-node garment prototype demonstrated real-time heatmap visualization at 99.2 Hz, proving the feasibility of high-rate distributed sensing on clothing.
Although the current prototype has not yet been fully optimized in terms of node size, power consumption, and mechanical robustness, it demonstrates the effectiveness of the proposed textile-based sensor networking technique. Its high refresh rate also provides sufficient communication headroom for future expansion.
Miniaturizing each node will reduce its thermal mass, enabling a faster transient response, while a lower-power circuit design will suppress self-heating and improve measurement accuracy.
This communication headroom can also be used to integrate a larger number of nodes, additional sensing modalities, or more robust packet-handling mechanisms.
These developments would support higher-resolution, more accurate, and more practical whole-body physiological monitoring using garment-integrated sensor networks.

\section*{Acknowledgment}
\footnotesize
This research was supported in part by JSPS KAKENHI JP25K03109 and JST CRONOS, Japan, Grant Number JPMJCS25N4. 
The authors used ChatGPT (OpenAI) for English language editing and Gemini (Google) for programming assistance in developing the visualization tool shown in Fig.~\ref{fig:heat_tracking}.

\end{document}